\documentclass[aps,showpacs,preprintnumbers,superscriptaddress,merge,notitlepage, nofootinbib]{revtex4-1}
\usepackage{hyperref,amsmath,subfigure,color}
\usepackage{graphicx}

\usepackage[compat=1.1.0]{tikz-feynman}
\usepackage{braket}

\begin{document}
\title{Superradiant pion clouds around primordial black holes} 
\date{\today}
\author{Paulo B. Ferraz}
\email{paulo.ferraz@fc.up.pt}
\affiliation{Departamento de F\'{\i}sica e Astronomia da Faculdade de Ci\^{e}ncias da Universidade do Porto, Rua do Campo Alegre 1021/1055, 4169-007 Porto, Portugal}
\affiliation{Departamento de F\'{\i}sica Te\'orica y del Cosmos, Universidad de Granada, Granada-18071, Spain}

\author{Thomas W. Kephart}
\email{thomas.w.kephart@vanderbilt.edu}
\affiliation{Department of Physics and Astronomy, Vanderbilt University, Nashvile, TN 37235, USA}

\author{Jo\~{a}o G.~Rosa}
\email{jgrosa@uc.pt}
\affiliation{Departamento de F\'{\i}sica da Universidade de Coimbra and CFisUC \\ Rua Larga, 3004-516 Coimbra, Portugal}
\affiliation{Departamento de F\'{\i}sica, Universidade de Aveiro, Campus de Santiago, 3810-183 Aveiro, Portugal}

%\preprint{[TO DO]}

%%%%%%%%%%%%%%%%%%%%%%%%%%%%%%%%%%%%%%%%%%%%%%%%%%%%%%%%%%%%%%%%%%%%%%%%%%%%%%%%%%%%%%%%%%%%%%%%%%%%%%%%%%%%%%%%%%%%%%%%%%%%%%%%%%%%%%%%
%%%%%%%%%%%%%%%%%%%%%%%%%%%%%%%%%%%%%%%%%%%%%%%%%%%%%%%%%%%%%%%%%%%%%%%%%%%%%%%%%%%%%%%%%%%%%%%%%%%%%%%%%%%%%%%%%%%%%%%%%%%%%%%%%%%%%%%%

\begin{abstract}
We show that highly spinning primordial black holes of mass $M\sim 10^{12}$ kg, potentially born in a matter-dominated era after inflation, can produce clouds of pions in their vicinity via the superradiant instability, with densities up to that of nuclear matter.  We discuss the electromagnetic signatures of this process, via neutral pion decay and charged pion annihilation into photons, computing in particular their contribution to the isotropic gamma-ray background. This allows us to place upper bounds on the abundance of such primordial black holes that are comparable to the ones obtained from Hawking evaporation. We also discuss the possibility of directly observing such clouds in high-redshift superclusters.
\end{abstract}

\maketitle

%\tableofcontents
%%%%%%%%%%%%%%%%%%%%%%%%%%%%%%%%%%%%%%%%%%%%%%%%%%%%%%%%%%%%%%%%%%%%%%%%%%%%%%%%%%%%%%%%%%%%%%%%%%%%%%%%%%%%%%%%%%%%%%%%%%%%%%%%%%%%%%%%
%%%%%%%%%%%%%%%%%%%%%%%%%%%%%%%%%%%%%%%%%%%%%%%%%%%%%%%%%%%%%%%%%%%%%%%%%%%%%%%%%%%%%%%%%%%%%%%%%%%%%%%%%%%%%%%%%%%%%%%%%%%%%%%%%%%%%%%%

\section{Introduction}

The superradiant instability of rotating black holes in general relativity has attracted a significant interest in the recent literature, given its potential to yield an astrophysical laboratory for  particle physics beyond the Standard Model \cite{Arvanitaki:2009fg, Arvanitaki:2010sy, Rosa:2011my, Pani:2012bp, Rosa:2012uz, Witek:2012tr, Brito:2013wya, Brito:2014wla, Arvanitaki:2014wva, Arvanitaki:2016qwi, Baryakhtar:2017ngi, Brito:2017wnc, Rosa:2017ury, Baumann:2018vus,  Hannuksela:2018izj, Isi:2018pzk, Boskovic:2018lkj, Ikeda:2019fvj, Ghosh:2018gaw, Berti:2019wnn, Baumann:2019eav, Sun:2019mqb}. Superradiant amplification of low-frequency wave modes   \cite{Zeldovich, Starobinsky, Teukolsky:1973, Press:1973zz, Teukolsky:1974yv} results from the dissipative nature of a black hole's event horizon, combined with rotation in the Kerr case, and it is the trapping of such modes in the black hole's vicinity that leads to continuous amplification and a consequent instability (see e.g.~\cite{review} for a recent review). In the case of massive fields, it is the black hole's gravitational potential that provides an intrinsic trapping mechanism \cite{Damour:1976, Zouros:1979iw, Detweiler:1980uk, Furuhashi:2004jk, Cardoso:2005vk, Dolan:2007mj, Rosa:2009ei, Dolan:2012yt, East:2017ovw, East:2017mrj, Dolan:2018dqv}, yielding a natural realisation of Press and Teukolsky's original ``black hole bomb" idea \cite{Press:1972zz, Cardoso:2004nk}. Since, to leading order, the gravitational potential falls as $1/r$ away from the black hole, massive fields can be trapped in Hydrogen-like bound states with an effective ``fine structure constant $\alpha_\mu = \mu G M/\hbar c$, where $\mu$ is the field's mass and $M$ is the mass of the black hole:
\begin{equation}
	E_n=\hbar \omega_n\simeq \mu c^2-{\alpha_\mu^2\mu c^2\over 2n^2}
	\label{Hspectrum}
\end{equation}
which is valid in the non-relativistic regime $\alpha_\mu \ll 1$, with $n=1,2, \ldots$ denoting the principal quantum number as for the Hydrogen atom. Superradiance occurs for $\omega_n< m\Omega$, where $m$ is the azimuthal angular momentum quantum number and $\Omega$ is the black hole's angular velocity at the event horizon. As we detail below, this condition translates into an upper bound on the dimensionless mass coupling, given the upper bound on the black hole's spin:
\begin{equation}
	\alpha_\mu \sim 0.1\left({\mu\over 10^{-11}\ \mathrm{eV}}\right)\left({M\over M_\odot}\right)<1/2~.
	\label{alpha_mu}
\end{equation}
We thus see that astrophysical black holes, with masses larger than the Sun, can only suffer from superradiant instabilities for ultra-light fields. Such fields must also carry an integer spin, given that superradiant amplification can be seen as a stimulated emission process, and only bosonic fields can have large occupation numbers in a given bound state. The literature has thus focused on ultra-light fields such as axions and axion-like particles, hidden photons or massive gravitons, predicted in some extensions of the Standard Model ands general relativity.

Astrophysical black holes may, however, not be the only black holes in the Universe, and there has also been an increasing interest in primordial black holes formed in the early Universe from the direct gravitational collapse of large overdensities in the primeval plasma \cite{Hawking:1971ei, Zeldovich-Novikov, Carr:1974nx}. Small-scale overdensities may arise in a plethora of cosmological scenarios, including certain inflationary models, curvaton models or cosmological phase transitions (through bubble collisions), and give rise to a broad mass spectrum for primordial black holes (see e.g.~\cite{Carr:2009jm} and references therein). It has been proposed that such black holes may constitute a significant fraction of the dark matter in the Universe and also that the binary black hole mergers observed recently by LIGO and Virgo may have a primordial origin (see e.g.~\cite{Carr:2016drx} and references therein). 

Given that primordial black holes may have much smaller masses than those formed through stellar collapse and other astrophysical mechanisms, it is thus natural to investigate whether superradiant instabilities can occur for heavier bosonic fields and have an astrophysical impact. This was considered, for instance, in \cite{Rosa:2017ury}, where it was shown that primordial black holes with $M\sim (10^{-7}-10^{-6})M_\odot$ can generate superradiant clouds of QCD axions in the preferred mass range where the latter can account for dark matter, $\mu\sim 10^{-5}$ eV, and that the stimulated axion decay into photons could be responsible for some of the brightest radio explosions in the cosmos, akin to the observed Fast Radio Bursts. 

Here we will be interested in investigating whether the best-known pseudo-scalar particles in nature - pions - can be produced through superradiance in the early Universe, and whether this has any observational impact. Since the pion mass is around $\sim 140$ MeV, superradiance can only occur for primordial black holes with $M\lesssim 10^{12}$ kg. There are already stringent constraints on the abundance of primordial black holes with mass around this upper bound, since they should be evaporating today and thus emit photons that contribute to the isotropic gamma-ray background (IGRB) \cite{Carr:2009jm, Arbey:2019vqx}. Pions potentially produced by superradiant instabilities around such black holes also contribute to the IGRB through their decay or annihilation into photons. This may then allow us to place new constraints on the abundance of such primordial black holes, depending not only on their mass but also on their spin\footnote{Note that Hawking evaporation also exhibits a mild dependence on the black hole spin (see e.g.~\cite{Arbey:2019vqx}), but superradiance is an intrinsic feature of spinning black holes.}. We will show, in particular, that an efficient pion production can only occur for highly spinning black holes, as the result of the pions' short lifetime, so that the IGRB directly constrains particular primordial formation mechanisms potentially leading to such large spins, such as an early matter-dominated epoch \cite{Harada:2017fjm}.

In addition to indirect constraints from the IGRB, we will also investigate the possibility of directly probing superradiant pion production by looking at high-redshift sources containing a significant number of highly spinning black holes. 

This work is organized as follows. In the next section we review the spectrum of superradiant bound states for massive scalar fields in the Kerr spacetime. In particular, we obtain a new analytical expression for the growth rate of the leading superradiant mode by fitting the results obtained by numerical solutions of the Klein-Gordon equation. In Sec. III, we discuss the development of superradiant instabilities for both neutral and charged pions, including in the dynamics the effects of neutral pion decay and charged pion annihilation into photons, as well as discussing the saturation of the pion clouds at around the nuclear density due to pion self-interactions. We then use these results to compute the contribution of these processes to the IGRB and use available observational data to place upper bounds on the primordial black hole abundance. In Sec. IV we discuss the potential for directly detecting gamma-rays from neutral pion decay and charged pion annihilation in primordial superradiant clouds. We summarize our main results and discuss future prospects in Sec. V. We consider the metric signature $(-,+,+,+)$ and, unless explicitly stated, we set $G=\hbar=c=1$.

\section{Superradiant instabilities for massive scalar fields in the Kerr spacetime}

A Kerr black hole is described by the line element, in Boyer-Lindquist coordinates (see e.g.~\cite{Townsend:1997ku}):
\begin{equation}
	ds^2=-\frac{\Delta-a^2\sin^2\theta}{\Sigma}dt^2-2a\sin^2\theta\frac{r^2+a^2-\Delta}{\Sigma}dtd\phi+\sin^2\theta\frac{(r^2+a^2)^2-a^2\Delta\sin^2\theta}{\Sigma}d\varphi^2+\frac{\Sigma}{\Delta}dr^2+\Sigma d\theta^2~,
	\label{lineelement}
\end{equation}
where
\begin{equation}
	\Delta = r^2+a^2-2Mr\;, \quad\quad \Sigma=r^2+a^2\cos^2\theta~.
\end{equation}
This geometry has coordinate singularities at $r_{\pm}=M\pm\sqrt{M^2-a^2}$, which correspond to the event horizon and the Cauchy horizon, respectively, and a curvature singularity at $r=0$, with an ergosurface at $r_e = M+\sqrt{M^2-a^2\cos^2\theta}$. The angular velocity of the black hole at the event horizon is given by:
\begin{equation}
	\Omega \equiv \frac{a}{r_+^2+a^2} = \frac{\tilde{a}}{2r_+}~,
\end{equation}
where $\tilde{a} = a/M$.

In the Kerr spacetime, the Klein-Gordon equation for a scalar field with mass $\mu$, $(\nabla_{\mu}\nabla^{\mu}-\mu^2)\phi=0$, can be separated into its angular and radial parts through a mode expansion of the form:
\begin{equation}
	\phi(t,r,\theta,\varphi)=\sum_{\omega, n, l, m}e^{-i\omega t +im\varphi}S_{lm}(\theta)R_{nlm}(r)~.
\end{equation}
This yields the equations:
\begin{align}
	\frac{1}{\sin\theta}\partial_{\theta}\big(\sin\theta\partial_{\theta}S\big)+\Big[a^2(\omega^2-\mu^2)\cos^2\theta-\frac{m^2}{\sin^2\theta}+\lambda\Big]S&=0~,
	\label{spheroidal}\\
	\Delta\partial_r\big(\partial_rR\big)-\Delta\Big[\mu^2r^2+a^2\omega^2-2\omega m a r+\big(\omega(r^2+a^2)-ma\big)^2+\lambda\Big]R&=0~,
	\label{radial}
\end{align}
where, for simplicity, we have dropped the labels $(n, l, m)$. The separation constant $\lambda$ corresponds to the eigenvalue of the angular spheroidal harmonic in Eq. (\ref{spheroidal}) and is given by the series expansion:
\begin{equation}
	\lambda = l(l+1)+\sum_{k=1}^{\infty}c_{klm}(aq)^{2k}~,
	\label{eigen}
\end{equation}
with $q=\sqrt{\mu^2-\omega^2}$, where the coefficients $c_{klm}$ can be found e.g.~in \cite{Abramowitz, Berti:2005gp}. In the non-relativistic regime, $\alpha_\mu\ll 1$, we have $aq\ll 1$ for the superradiant bound states, as we will see below, and thus we may approximate $\lambda \simeq l(l+1)$.

Following the procedure described in \cite{Rosa:2012uz}, it is useful to write the radial equation in terms of the dimensionless coordinate $x=(r-r_+)/r_+$, resulting in:
\begin{equation}
	x^2(x+\tau)^2\partial_x^2R+x(x+\tau)(2x+\tau)\partial_xR+V(x)R=0~,
	\label{dimequa}
\end{equation}
where
\begin{equation}
	V(x)=\big[x(x+\tau)\overline{\omega}+(2-\tau)(\overline{\omega}-m\overline{\Omega})\big]^2+x(x+\tau)\big[(\tau-1)\overline{\omega}^2+2(2-\tau)\overline{\omega}m\overline{\Omega}-\overline{\mu}^2(x+1)^2-\lambda\big]~.
\end{equation}
In the above expression, barred quantities are dimensionless and are defined as $\overline{\alpha}=r_+\alpha$, and $\tau = (r_+-r_-)/r_+$, such that $\tau=0$ corresponds to an extremal  Kerr black hole ($\tilde{a}=1$) and $\tau = 1$ corresponds to a non-rotating Schwarzschild black hole. 

Although the radial equation (\ref{radial}) does not have an exact analytical solution, in the non-relativistic regime $\alpha_\mu\ll 1$ (equivalently $\overline{\mu}\ll 1$) we can use a procedure \cite{Starobinsky} that consists in separating the exterior of the black hole into two overlapping regions: a near-horizon region, $\overline{\omega}x\ll l$, and a far-region, $x\gg 1$. The radial equation can be solved analytically in both regions, and we can match the two solutions in the overlapping region $1\ll x \ll l/\overline{\omega}$. Since for quasi-bound states $\omega \simeq \mu$ to leading order, the validity of this procedure is justified in the small mass coupling limit.

In the near-region, we have:
\begin{equation}
	x^2(x+\tau)^2\partial_x^2R+x(x+\tau)(2x+\tau)\partial_xR+V(x)R=0\;,\quad\quad V(x)\simeq\big[(2-\tau)(\overline{\omega}-m\overline{\Omega})\big]^2-x(x+\tau)\lambda~,
	\label{nearregion}
\end{equation}
where we defined $\overline{\varpi}=(2-\tau)(\overline{\omega}-m\overline{\Omega})$. This has analytical solutions that can be expressed in terms of hypergeometric functions, and the physical solution with ingoing boundary conditions at the horizon at $x=0$ is of the form:
\begin{equation}
	R_{near}(x)=A\Big(\frac{x}{x+\tau}\Big)^{-i\overline{\varpi}/\tau}{}_2F_1(l+1,-l,1-2i\overline{\varpi}/\tau,-x/\tau)~.
	\label{near}
\end{equation}
Using the asymptotic properties of the hypergeometric function \cite{Abramowitz}, taking the limit $x\gg\tau$ of the latter, we obtain:
\begin{equation}
	R_{near}(x)\simeq A\Gamma(1-2i\overline{\varpi}/\tau)\left[\frac{\Gamma(2l+1)}{\Gamma(l+1)\Gamma(l+1-2i\overline{\varpi}/\tau)}\left(\frac{x}{\tau}\right)^l+\frac{\Gamma(-2l-1)}{\Gamma(-l)\Gamma(-l-2i\overline{\varpi}/\tau)}\left(\frac{x}{\tau}\right)^{-l-1}\right]~.
\end{equation}

In the far-region, Eq. (\ref{dimequa}) reduces to:
\begin{equation}
	x^2\partial_x^2R+2x\partial_xR+(-\overline{q}^2x^2+2\overline{q}\nu x-\lambda)R=0~,
\end{equation}
where
\begin{equation}
	\nu = \left(\frac{2-\tau}{2}\right)\left(\frac{\overline{\omega}-\overline{q}}{\overline{q}}\right)~.
\end{equation}
The solutions can be expressed in terms of confluent hypergeometric functions, and the bound state solutions that decay exponentially at infinity are of the form:
\begin{equation}
	R_{far}(x)=Bx^le^{-\overline{q}x}U(l+1-\nu,2l+2,2\overline{q}x)~.
\end{equation}
Taking the limit $\overline{q}x\ll 1$ \cite{Abramowitz}, we then find:
\begin{equation}
	R_{far}(x)\simeq B\frac{\pi}{\sin((2l+2)\pi)}\left[\frac{x^l}{\Gamma(-l-\nu)\Gamma(2l+2)}-(2\overline{q})^{-(2l+1)}\frac{x^{-l-1}}{\Gamma(l+1-\nu)\Gamma(-2l)}\right]\;.
\end{equation}
As expected, the near-region and far-region solutions have the same behaviour in their common domain of validity and one can obtain the spectrum of the quasi-bound states by matching the coefficients of $x^l$ and $x^{-l-1}$, which gives the condition:
\begin{equation}
	\frac{\Gamma(-l-\nu)\Gamma(2l+2)}{\Gamma(l+1-\nu)\Gamma(-2l)}=-(2\overline{q}\tau)^{2l+1}\frac{\Gamma(-2l-1)\Gamma(l+1)\Gamma(l+1-2i\overline{\varpi}/\tau)}{\Gamma(-l)\Gamma(2l+1)\Gamma(-l-2i\overline{\varpi}/\tau)}~.
	\label{matching}
\end{equation}
For bound states in the small mass coupling (non-relativistic) limit, $\overline{q}\ll 1$, and to leading order the right-hand side of Eq.~(\ref{matching}) vanishes, yielding the condition:
\begin{equation}
	\frac{1}{\Gamma(l+1-\nu^{(0)})} = 0 \implies l+1-\nu^{(0)} = -n_r~,
\end{equation}
where $n_r$ is a non-negative integer, corresponding to the number of radial nodes. From the definition of $\nu$, we may solve for the bound state frequency using the approximation $\overline{\omega}=\overline{\omega}^{(0)}+\delta\overline{\omega}$, which results in the Hydrogen-like spectrum anticipated in Eq.~(\ref{Hspectrum}):
\begin{equation}
	\overline{\omega}^{(0)}\simeq \overline{\mu}\;,\quad\quad \delta\overline{\omega}=-\overline{\mu}\frac{(\mu M)^2}{2(l+1+n_r)^2}\equiv -\overline{\mu}\frac{(\mu M)^2}{2n^2}\;,
\end{equation}
where $n=l+1+n_r$. To determine whether a mode is superradiant and hence stable or not, we need to compute the sub-leading imaginary part of the bound state frequency. To do so, we expand the left-hand side of the matching condition with $\nu=\nu^{(0)}+\delta\nu$ and evaluate the right-hand side with the leading order result $\omega^{(0)}$. To cancel the poles in the gamma functions, we use the following results:
\begin{equation}
	\lim_{z\rightarrow-n}\frac{\psi(z)}{\Gamma(z)}=(-1)^{n+1}n!~,\quad\quad \frac{\Gamma(-2l-1)}{\Gamma(-l)}=\frac{(-1)^{l+1}}{2}\frac{l!}{(2l+1)!}~,\quad\quad \frac{\Gamma(l+1-x)}{\Gamma(-l-x)}=(-1)^lx\prod_{k=1}^{l}(k^2-x^2)~,
\end{equation}
where $\psi(z)=\Gamma'(z)/\Gamma(z)$. After some algebra, we obtain:
\begin{equation}
	\omega_IM=-\frac{1}{2}A_{nl}\left(\frac{\varpi M}{\tau}\right)(\mu M)^{4l+5}\left(\frac{r_+-r_-}{r_++r_-}\right)^{2l+1}~,
	\label{imaginarypart}
\end{equation}
with 
\begin{equation}
	A_{nl}=\Bigg(\frac{l!}{(2l+1)!(2l)!}\Bigg)^2\frac{(l+n)!}{(n-l-1)!}\frac{4^{2l+2}}{n^{2l+4}}\prod_{k=1}^l\Bigg(k^2+16\Big(\frac{\varpi M}{\tau}\Big)^2\Bigg)~.
	\label{coeffA}
\end{equation}
Hence, a mode is unstable for $\omega_I<0$, which yields the superradiance condition $\varpi < 0$ or equivalently: 
\begin{equation}
	 \omega_R<m\Omega~.
	\label{supercondition}
\end{equation}
Since, to leading order, $\omega_R\simeq \mu$, in the non-relativistic regime, this implies $\alpha_\mu<m/2$ for the dimensionless mass coupling, and $\alpha_\mu<1/2$ for the fastest growing mode with $l=m=1$ as anticipated above in Eq.~(\ref{alpha_mu}).

Beyond the small mass coupling limit, numerical methods must be employed to compute the spectrum of quasi-bound states. Different methods have been used in the literature  (see e.g.~\cite{Dolan:2007mj, Rosa:2011my, Rosa:2012uz}), and here we focus on a simple but effective shooting method where one integrates the radial equation from the horizon up to a large distance, where one then minimizes the solution in the complex frequency plain to look for quasi-bound states. Boundary conditions at the horizon can be obtained by writing the ingoing solution as a series:
\begin{equation}
	R(x)=x^{-i\overline{\varpi}/\tau}\sum^{\infty}_{n=0}a_nx^n~,
	\label{rtaylor}
\end{equation}
where the $a_n$ coefficients can be computed by substituting this \textit{ansatz} into Eq. (\ref{dimequa}) and making use of a symbolic algebraic package such as Mathematica to solve it order by order. One may set $a_0=1$ since the normalisation is not relevant in computing the quasi-bound state spectrum.

\begin{figure}[htbp]
	\centering
	\includegraphics[scale=0.5]{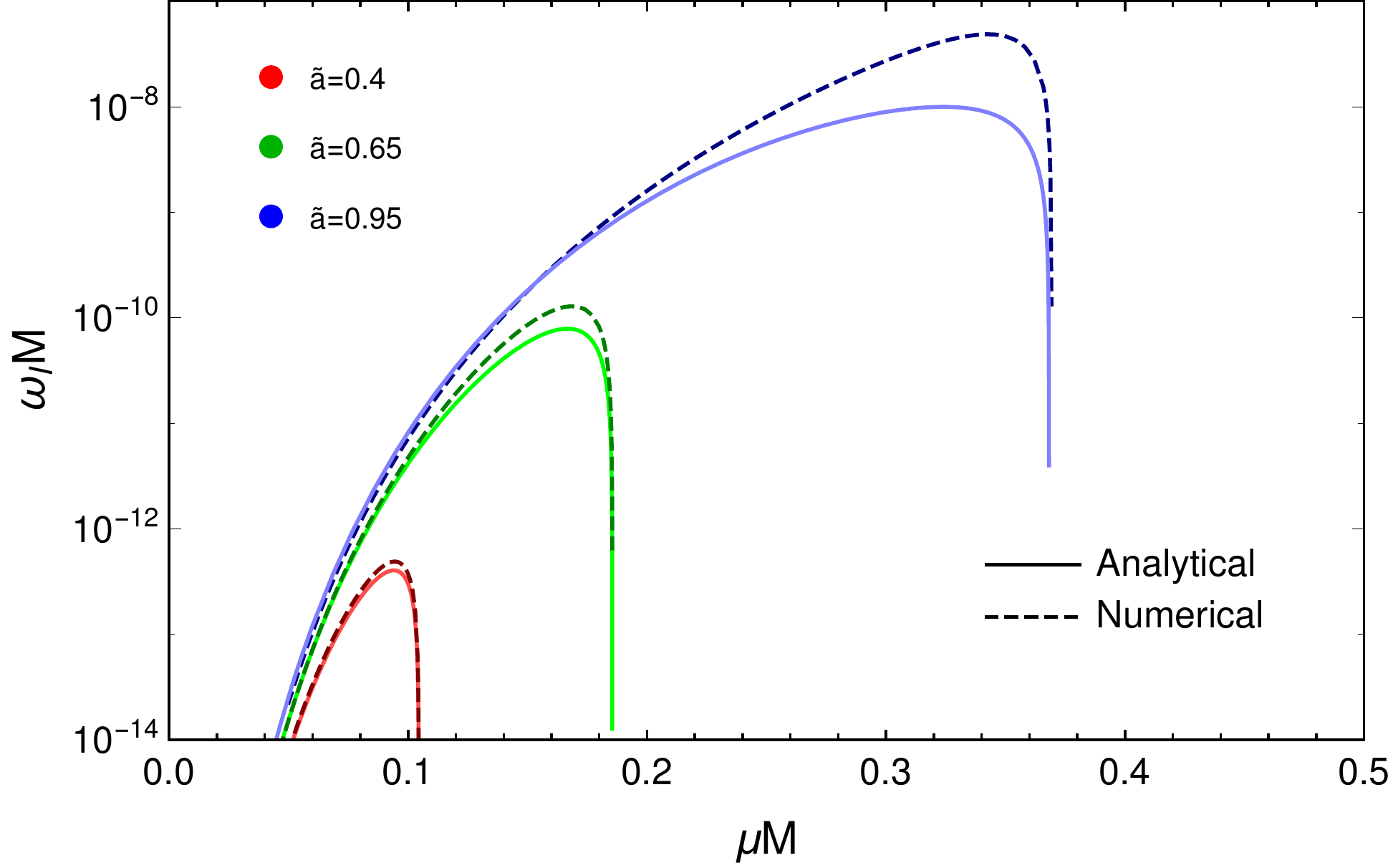}
	\caption{Imaginary part of the bound state frequency for the fastest superradiant mode $l=m=1$, $n=2$, alongside the approximate analytical prediction, as a function of the dimensionless mass coupling $\alpha_\mu=\mu M$ and for different values of the black hole spin.}
	\label{semajuste}
\end{figure}

In Fig.~\ref{semajuste} we show the numerical results for the imaginary part of the leading superradiant mode, $l=m=1$, $n=2$, as a function of the dimensionless coupling $\mu M$ and for different values of the black hole spin $\tilde{a}$.

\begin{figure}[htbp]
	\centering
	\includegraphics[scale=0.5]{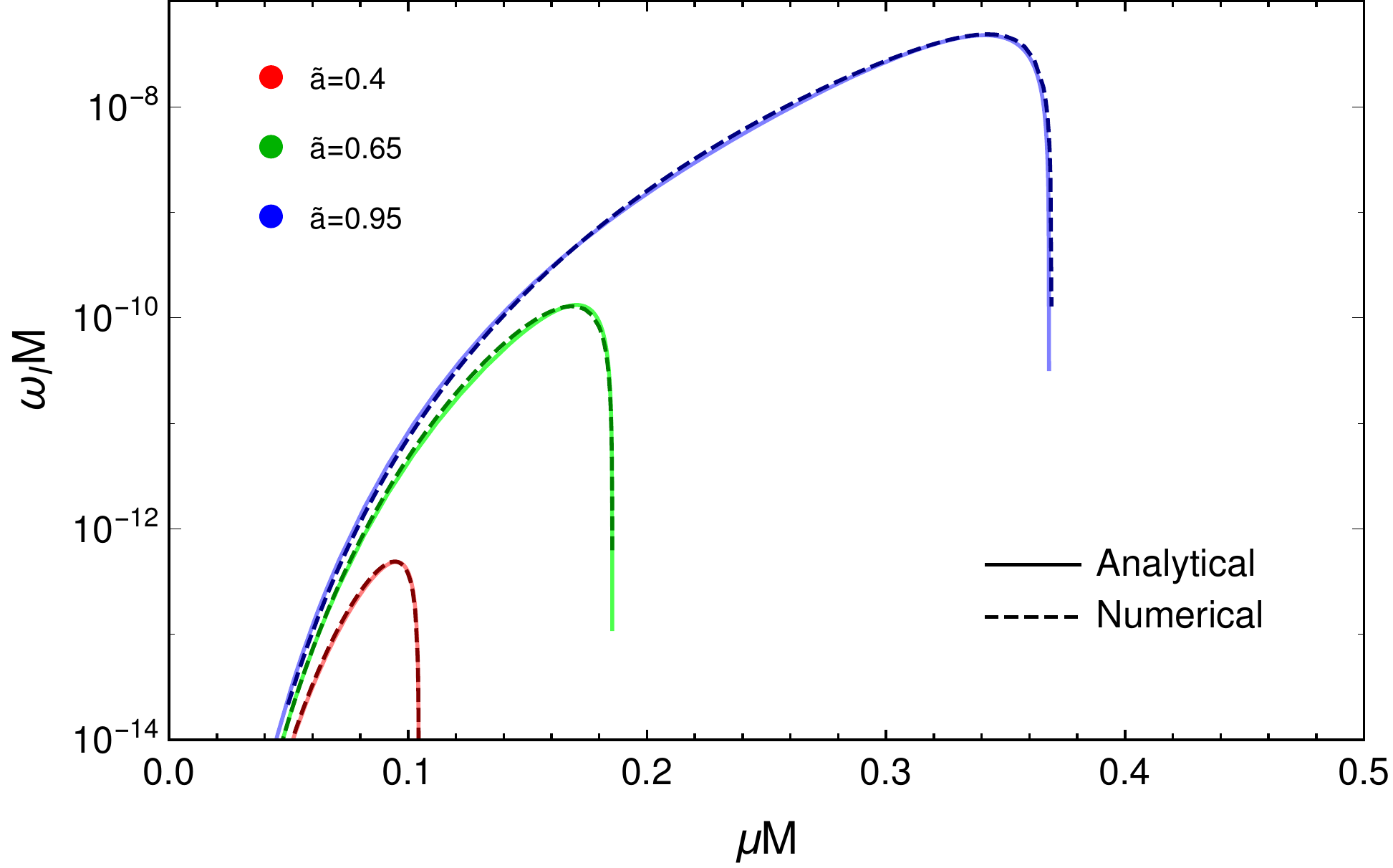}
	\caption{Imaginary part of the bound state frequency for the fastest superradiant mode, compared to the corrected analytical expression, as a function of the dimensionless mass coupling $\alpha_\mu=\mu M$ for different values of the black hole spin.}
	\label{comajuste}
\end{figure}

As one can easily see, the analytical expression derived in the non-relativistic limit deviates considerably from the numerical results for $\alpha_\mu\gtrsim 0.1$, and for large black hole spins this is the regime where the superradiant instability develops faster. As we will discuss in the next section, this is, in fact, the regime where pions can be produced efficiently in superradiant clouds, so it will be useful to find a better analytical description of the growth rate. Since the deviations from the analytical result are more pronounced close to the endpoint of the superradiant instability, we have looked for correction factors of the form $1+f(\tilde{a})(\mu/ m\Omega)^p$. We have found that factors of this form with $p=6$ yield a good approximation to the numerical results for the leading superradiant mode, and fitted the coefficient $f(\tilde{a})$ to the numerical data for different black hole spins. In the end, we find a corrected expression of the form:
\begin{equation}
	(\omega_IM)_{corrected}=-\frac{1}{2}A_{2,1}\left(\frac{\varpi M}{\tau}\right)(\mu M)^9\left(\frac{r_+-r_-}{r_++r_-}\right)^3\left(1+{3\over2}\frac{\tilde{a}^{9/5}}{\sqrt{1-\tilde{a}}}\left(\frac{\mu}{\Omega }\right)^6\right)~,
	\label{imaginarypartadjusted}
\end{equation}
where $A_{2,1}$ is given by Eq. (\ref{coeffA}) for $l=1$ and $n=2$. The corrected imaginary part of the frequency is depicted in Fig.~\ref{comajuste}, where we clearly see the very good agreement with the numerical results. We find deviations of $\lesssim 10\%$ between the numerical data and the corrected expression for most black hole spins (and $\lesssim 50\%$ close to the extremality), which is sufficiently good for our purposes.

Since the superradiant quasi-bound state $n=2$, $l=m=1$ is populated exponentially faster than all others, we will focus on this state in our discussion henceforth. The occupation number in this ``2p"-cloud in Hydrogen-like notation grows as $e^{\Gamma_s t}$, where $\Gamma_s=2\omega_I$. This cloud has approximately a toroidal shape with radii $\langle r \rangle = 5r_0$ and $\Delta r =\sqrt{\langle r^2\rangle -\langle r\rangle^2} =\sqrt{5}r_0$, where the ``Bohr radius'' is given by:
\begin{equation}
	r_0={1\over \mu\alpha_\mu}=\frac{\alpha_{\mu}^{-2}}{1+\sqrt{1-\tilde{a}^2}}r_+\;.
\end{equation}
Thus, in the non-relativistic regime, the cloud is localized far away from the horizon and gravitational effects may be neglected in our subsequent study of particle decay and annihilation. Note also that the r.m.s.~velocity of the particles, $\sqrt{\langle v^2\rangle}\simeq (\alpha_{\mu}/2)$, is non-relativistic for $\alpha_{\mu}\lesssim1$. In the next section, we will then study the dynamics and phenomenology associated with both neutral and charged pions produced by superradiant instabilities with these assumptions.

%%%%%%%%%%%%%%%%%%%%%%%%%%%%%%%%%%%%%%%%%%%%%%%%%%%%%%%%%%%%%%%%%%%%%%%%%%%%%%%%%%%%%%%%%%
%%%%%%%%%%%%%%%%%%%%%%%%%%%%%%%%%%%%%%%%%%%%%%%%%%%%%%%%%%%%%%%%%%%%%%%%%%%%%%%%%%%%%%%%%%
%%%%%%%%%%%%%%%%%%%%%%%%%%%%%%%%%%%%%%%%%%%%%%%%%%%%%%%%%%%%%%%%%%%%%%%%%%%%%%%%%%%%%%%%%%

\section{Superradiant pion clouds}

\subsection{Neutral pions}

\begin{figure}[htbp]
	\centering
	\includegraphics[scale=0.4]{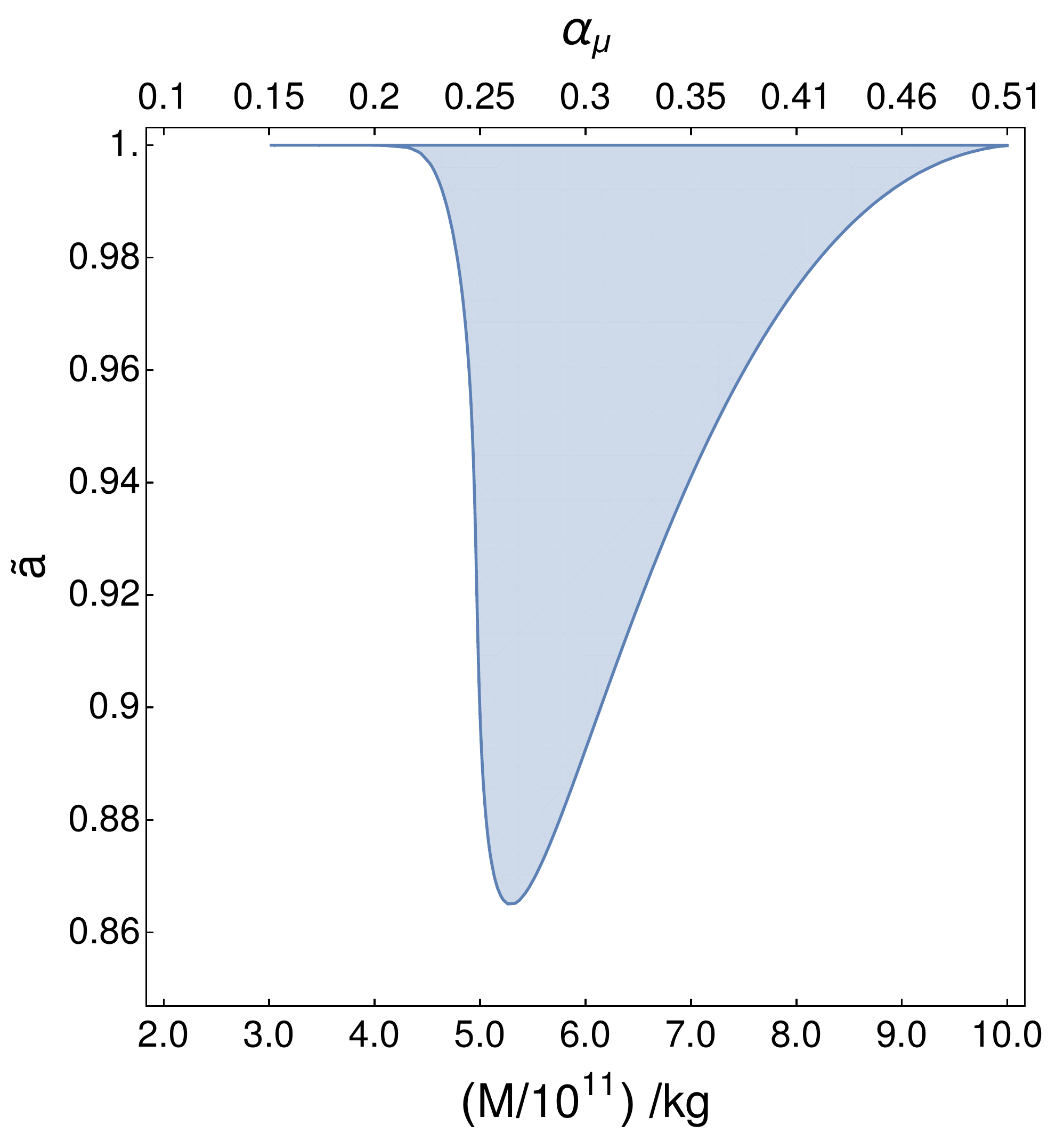}
	\caption{Region in the black hole mass-spin Regge plot in which an efficient neutral pion production through superradiance is possible.}
	\label{neutraleffect}
\end{figure}

Neutral pions are spin-0 particles with mass $\mu_0\simeq 135$ MeV and a mean lifetime $\Gamma_{\pi_0}^{-1}\simeq 8.4\times 10^{-17}$ s, decaying mostly into photon pairs $\pi^0\rightarrow\gamma\gamma$ \cite{Tanabashi:2018oca}. Neutral pion production around primordial black holes will thus be efficient if the superradiant instability develops faster than the pions decay, which leads to the condition $\Gamma_{eff}=\Gamma_s-\Gamma_{\pi_0}>0$. This is illustrated in Fig.~\ref{neutraleffect} in the black hole mass-spin Regge plot, where one can see that an efficient neutral pion production is possible for highly spinning black holes in the mass range $M\sim 10^{11}-10^{12}$ kg.

Black holes in this mass range may be formed from the collapse of large overdensities in the early Universe after inflation, when the corresponding modes re-enter the horizon. In conventional scenarios, the Universe is dominated by radiation following reheating after inflation, and such black holes would form at temperatures in the range $10^8-10^9$ GeV. In this case, the gravitational collapse of large overdensities is mostly spherical and the resulting black holes have been shown to have small spins at the percent level \cite{Chiba:2017rvs, Mirbabayi:2019uph, DeLuca:2019buf}. However, for lower values of the reheat temperature, the Universe should still be dominated by the inflaton field at the time of black hole formation in the mass range of interest. This corresponds to an early matter-dominated era, since the inflaton field behaves as non-relativistic matter while oscillating about the minimum of its potential\footnote{Assuming the inflaton field has a non-vanishing mass at the minimum.}. Since the background fluid is essentially pressureless, anisotropies can develop during the collapse and generate angular momentum in the process. It has been argued that the resulting centrifugal force may, in fact, prevent collapse, and that the few black holes that form have very large spins, including near-extremal black holes \cite{Harada:2017fjm}. Hence, in such a case it will be natural for neutral pion clouds to develop through superradiance around such black holes, and potentially there may be other mechanisms to form black holes with large natal spins.

One must take into account, however, that the superradiant instability cannot produce an arbitrarily large pion occupation number in the cloud, since pions are strongly interacting. In the effective pion Lagrangian, the leading self-interactions correspond to a quartic term in the pion potential $\lambda \phi^4/4!$, where $\lambda =\mu_0^2/2f_{\pi}^2\simeq 1 $, where the pion decay constant $f_\pi\simeq 93$ MeV. Such self-interactions should become important when the field value within the cloud reaches a critical value $\phi_{c0}\simeq \sqrt{12\mu_0^2/\lambda}$ for which the quadratic and quartic terms in the pion scalar potential become comparable. It has been shown that non-linear effects quench the superradiant instability, leading to the collapse of the cloud in ``bosenova"-like explosions \cite{Arvanitaki:2009fg, Yoshino:2012kn, Yoshino:2015nsa} similar to those observed in cold atom Bose-Einstein condensates \cite{bosenova}. Despite the complex dynamics of this process, the leading effect is a reabsorption of particles in the cloud by the black hole, as non-linearities transfer particle number from superradiant to non-superradiant states, although a small fraction of the particles may escape the cloud. Once the cloud occupation numbers are depleted by the bosenova-collapse and the field decreases below the critical value, the superradiant instability may restart, until a new bosenova event is triggered. Since such processes are fast, we may to a first approximation consider that the pions remain, on average, close to the critical occupation number once it is first reached. This may be computed by using the critical pion number density $n_{c0}\sim\mu_0\phi_c^2$ within the toroidal 2p-cloud as described above, yielding:
\begin{equation}
	N_{c0}\simeq 600\pi^2\alpha_{\mu0}^{-3}~,
\end{equation}
which depends on the black hole mass. For instance, for a black hole mass $M=5.5\times 10^{11}$ kg, we find $N_{c0}\simeq 3\times 10^5$, corresponding to a cloud density $\rho_{\phi}\simeq9\times 10^{17}\; \text{kg\,m}^{-3}$ that is close to the nuclear density $\rho_n\simeq 1.7\times 10^{18}\;\text{kg m}^{-3}$ ($1$ nucleon per $1 \text{fm}^3$), as one would expect since pion self-interactions reflect the underlying strong nuclear force.

We are interested in computing the photon flux associated with neutral pion decay in such critical superradiant clouds. First, let us note that the variation of the black hole's mass and spin are  $\Delta M = \mu_0 N_0$, and spin, $\Delta J = N_0$ for $N_0$ neutral pions produced in the $2p$-cloud, yielding the relation:
\begin{equation}
	\frac{\Delta M}{M} = \alpha_{\mu_0}\tilde{a}\frac{\Delta J}{J}~,
	\label{deltas}
\end{equation} 
where $M$ and $J$ denote the black hole's initial mass and spin. We may use this relation and the fact that the final values of the black hole's mass and spin must lie along the curve $\Gamma_{eff}=0$, corresponding to the boundary of the blue region in Fig.~\ref{neutraleffect}, to determine how much mass and spin a given black hole loses due to superradiance. The rate at which the cloud loses energy corresponds to the photon luminosity  $L_0\simeq 2N_{c0}\Gamma_{\pi_0}E_{\gamma}$, where $E_\gamma\simeq \mu_0/2$ since the pion's are non-relativistic. Hence, the lifetime of a critical superradiant neutral pion cloud is $t_{cloud}= \Delta M/L_0$. For instance, for a near-extremal black hole $\tilde{a}=0.99$ with mass $M=5.5\times 10^{11}$ kg, this yields $t_{cloud}\simeq 1.4$ Gyr, i.e.~pion superradiance can be active on cosmological time scales. 

Since black holes in the relevant mass range are evaporating today, $t_{ev}\sim 10$ Gyr, we conclude that superradiance shuts down well before the black hole's evaporate significantly, and we may, to a first approximation, neglect the effects of Hawking evaporation on the black hole's mass and spin evolution\footnote{Note that a significant loss of mass and spin through Hawking evaporation only occurs towards  the end of a black hole's lifetime \cite{Arbey:2019vqx}.}.

\begin{figure}[htbp]
	\centering
	\includegraphics[scale=0.5]{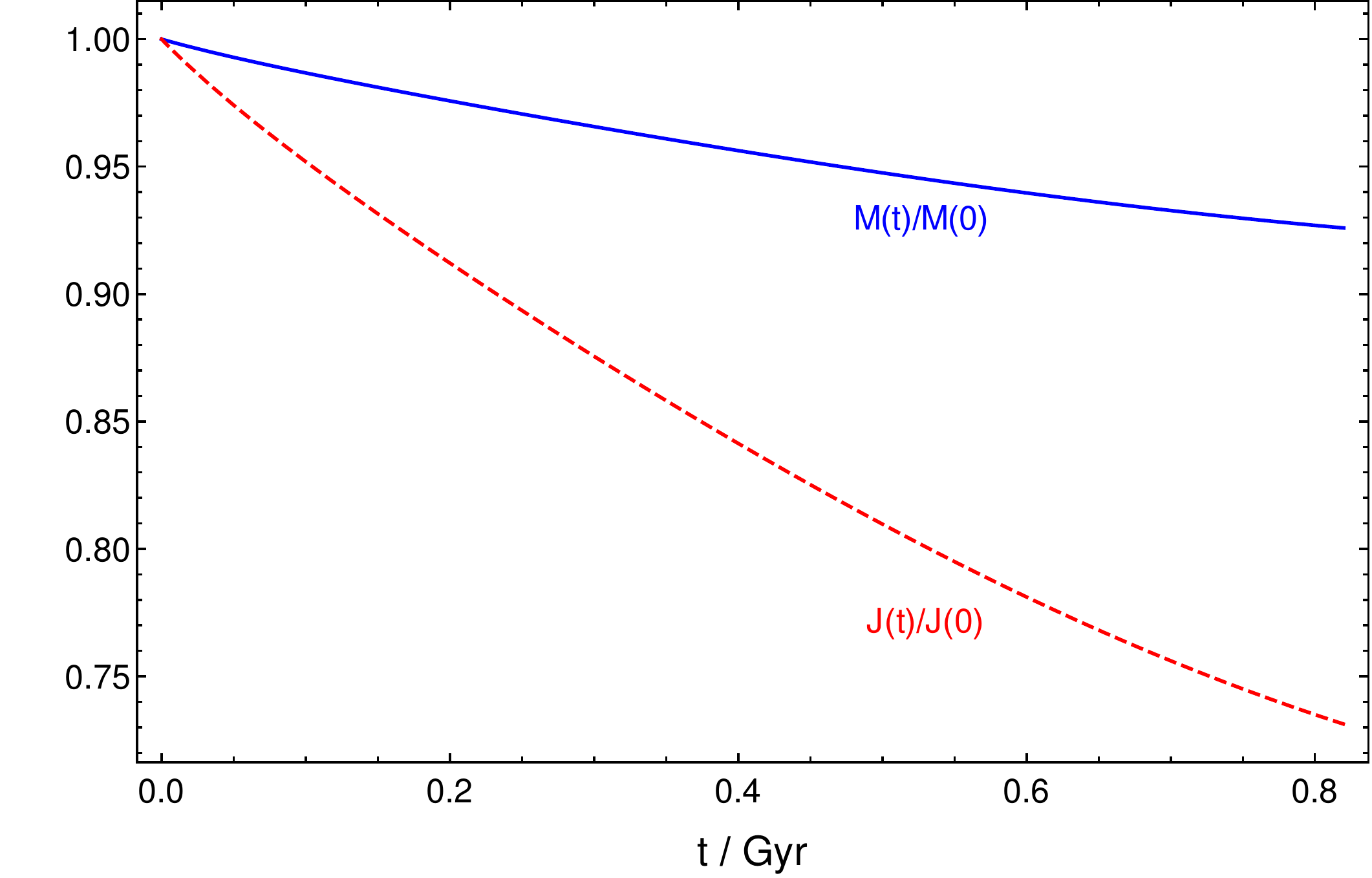}
	\caption{Dynamics of the black hole's mass and spin, normalized to their initial values $M(t=0)=5.5\times 10^{11}$ kg and $\tilde{a}(t=0)=0.99$ under neutral pion superradiant instabilities.}
	\label{neutroevo}
\end{figure}

We can do a more accurate computation by taking into account that the critical neutral pion number depends on the mass coupling $\alpha_{\mu0}$, which decreases as the black hole loses mass through superradiant pion production. Since the superradiant instability grows much faster than the cosmological time scale in which the black hole effectively loses mass and spin, we may assume that the cloud is always, on average, close to criticality, with the critical number varying adiabatically. While in the superradiant regime, the differential equations governing the dynamics of the black hole's mass and spin, as well as the number of photons emitted through neutral pion decay are then given by:
\begin{equation}
	\frac{dM}{dt} = -\mu_0 \Gamma_{s0}N_{c0}~,\quad
	\frac{dJ}{dt}=-\Gamma_{s0}N_{c0}~,\quad
	\frac{dN_{\gamma}}{dt}=2\Gamma_{\pi_0}N_{c0}~,
	\label{dynamics}
\end{equation}
where $N_{c0}= N_{c0}(M)$ and is thus time-dependent. Note that here we neglect stimulated decay and photon annihilation as considered e.g.~\cite{Rosa:2017ury} in the axion case, since such processes are sub-leading for pions, which are not produced in such large numbers as axions.  Once the system leaves the superradiant regime, $\Gamma_{eff}<0 $, all pions essentially decay into photons. Solving these equations numerically for the initial black hole mass and spin given above, we obtain that the production of neutral pions is effective over a period $t_{cloud} \simeq 0.8$ Gyr, slightly smaller than our earlier estimate given the increase in the cloud's luminosity over time as a result of the growth of the critical pion number with the decrease in $M$. In Fig.~\ref{neutroevo}, we show the time evolution of the black hole's mass and spin, normalized to their initial values, and we see that the black hole loses spin faster than it loses mass, as predicted above in Eq. (\ref{deltas}). 

Having determined the cosmological period during which superradiant pion production is active and the corresponding photon luminosity, we may now compute the contribution of this process to the isotropic gamma-ray background (IGRB) as a function of the primordial black hole abundance in the relevant mass and spin range. This diffuse background fills the intergalactic medium, corresponding to the photon flux remaining after subtracting the flux from known sources, and has been measured by different experiments, in particular HEAO-1 \cite{Gruber:1999yr}, COMPTEL \cite{Weidenspointner}, EGRET \cite{Strong:2004ry} and more recently FERMI-LAT \cite{Abdo:2010nz}.

Let us assume that the distribution of primordial black holes in the relevant mass-spin range is isotropic on large scales, thus originating also an isotropic photon flux from the decay of the neutral pions produced through the superradiant instability. The photon flux measured today will thus be a superposition of all the photons emitted at different epochs. We note that, even though our primordial black holes could form before reheating and the onset of radiation-domination, and that superradiant instabilities can thus be active from very early times, only photons emitted after recombination, in the cold dark matter (CDM) epoch, contribute to the IGRB, since before that emitted photons quickly thermalize with cosmic photons and other particles in the ambient plasma, and thus contribute instead to the Cosmic Microwave Background.

The emission rate per unit volume for a given energy $E$ at a cosmological time $t$ is then:
\begin{equation}
	\frac{dn_{\gamma}}{dt}(E,t)=n_{PBH}(t)\frac{dN_{\gamma}}{dt}(E,t)~,
\end{equation}
where $n_{\gamma}$ and $n_{PBH}$ are the number density of photons and primordial black holes, respectively. Since the Universe is expanding, both the number density of primordial black holes and the photons' energy at cosmological time $t$ will be observed today with redshift factors $(1+z)^{-3}$ and  $(1+z)^{-1}$, respectively. Writing the primordial black holes number density as $n_{PBH}(t) = \Omega_{PBH,0}(\rho_{c,0}/M(t))(1+z)^3$, where $\rho_{c,0}$ is the present value of the critical density, $\Omega_{PBH,0}$ is the abundance of primordial black holes today and $M(t)$ is the black hole mass, and following the same procedure as in \cite{Carr:2009jm, Arbey:2019vqx}, we estimate that the present flux of photons of energy $E$ is given by:
\begin{align}
	I\simeq& \frac{dI}{dE}E\nonumber\\
	=&\frac{1}{4\pi}\rho_{c,0}\Omega_{PBH,0}E_0\int^{t_{cloud}}_{t_{rec}}dt\frac{(1+z)}{M(t)}\frac{d^2N_{\gamma}}{dEdt}(E_0(1+z))~,
	\label{flux}
\end{align}
where $E_0$ is the photon energy measured today, $t_{rec}\simeq4\times 10^5$ yr is the time of recombination and $t_{cloud}\sim 1$ Gyr is the time at which superradiant pion production becomes ineffective as computed above. It is then clear that all relevant photons are emitted in the CDM-dominated era, for which $(1+z)=(t/t_0)^{-2/3}$. Note that here we are assuming that no late time mergers (see below) spin up the black holes and that only superradiance is active.

We will consider two approximations to the spectrum of photons from neutral pion decay. As a first approximation, we consider a monochromatic spectrum with $E_\gamma=\mu_0/2$, such that:
\begin{equation}
	\frac{d^2N_{\gamma}}{dEdt} = 2N_c\Gamma_{\pi_0}\delta(E-E_\gamma)~.
\end{equation}
This would be exact if the pions in the superradiant cloud were at rest, which is not truly realistic since they have a small average velocity $\sim \alpha_{\mu0}/2$ as mentioned earlier, such that photons are emitted with energies in a range bounded by $\mu_0/2(1\pm \alpha_{\mu0}/2)$. A better approximation is then to consider a Gaussian spectrum centered at half the pion's mass and width $\sigma = E_{\gamma}\alpha_{\mu0}/2$. The monochromatic approximation has the advantage of allowing for an analytical expression for the present photon flux: 
\begin{equation}
	I(E_0)=\frac{3}{4\pi}\rho_{c,0}\Omega_{PBH,0}\Gamma_{\pi_0}t_0\left(\frac{E_0}{E_{\gamma}}\right)^{3/2}\left.{N_{c0}\over M}\right|_{t=t_{e}(E_0)}~,
\end{equation}
where  the ratio $N_{c0}/M$ is evaluated at the time $t_{rec}<t_{e}< t_{cloud}$ at which photons of energy $E_0$ were emitted, such that $E_0(t_{e}/t_0)^{-2/3}=E_{\gamma}$. The time interval from recombination until the neutral pion cloud decays away thus sets the energy range of the present photon spectrum. For the Gaussian spectrum, the integral in Eq.~(\ref{flux}) has to be computed numerically.

\begin{figure}[htbp]
	\centering
	\includegraphics[scale=0.47]{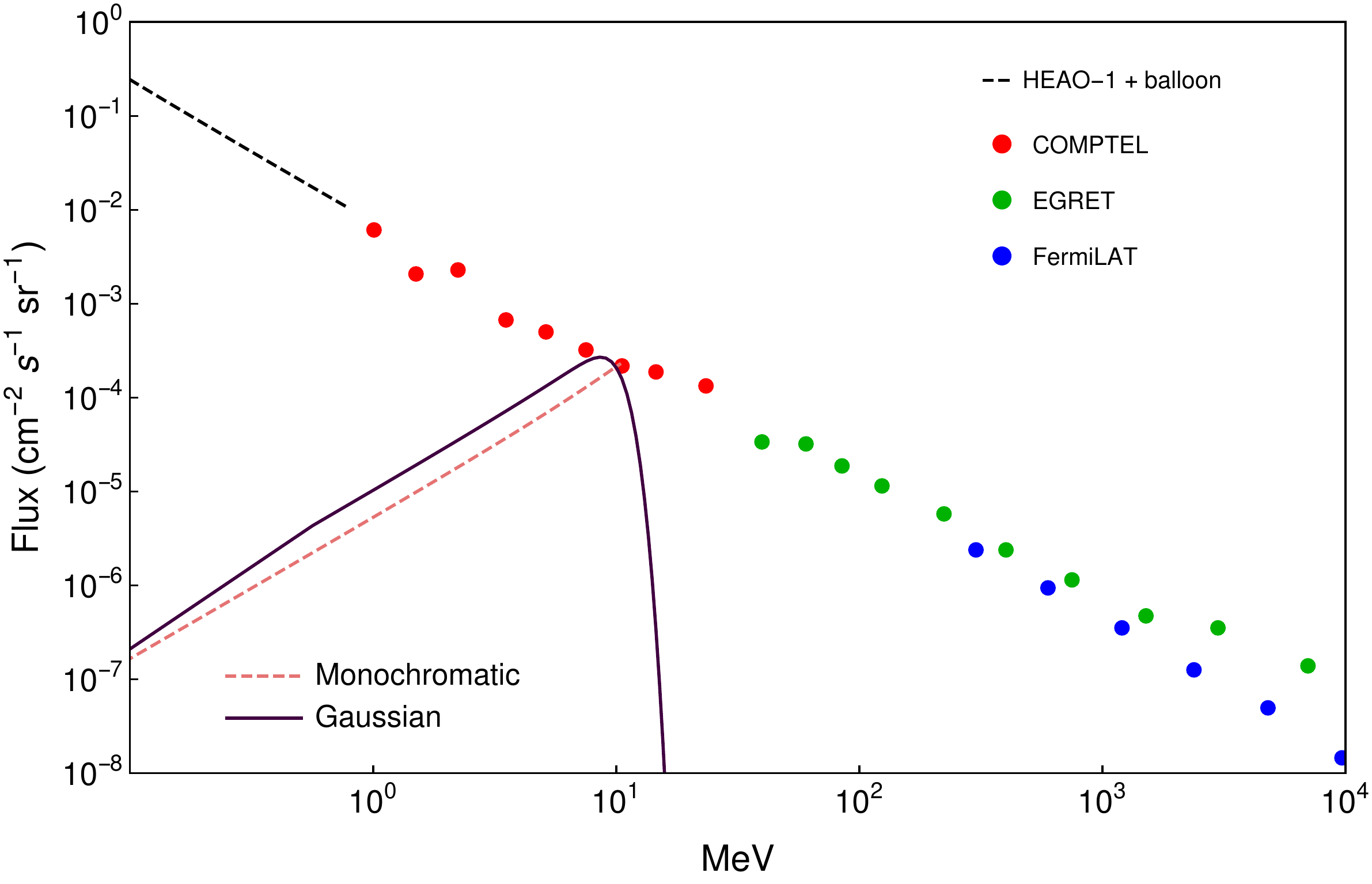}
	\caption{Photon flux from neutral pion decay in superradiant clouds, for a monochromatic (dashed curve) and Gaussian (solid curve) emission spectrum, considering a population of near-extremal primordial black holes ($\tilde{a}=0.99$) with mass $M=5.5\times10^{11}$ kg. Data points correspond to IGRB measurements by the quoted experiments.}
	\label{photonflux}
\end{figure}

In Fig.~\ref{photonflux} we show the prediction for the photon flux for both spectral approximations for a black hole with initial mass $M=5.5\times 10^{11}$ kg and initial spin $\tilde{a}=0.99$, alongside the measured IGRB flux. In this figure we have selected the maximum value of the fraction of dark matter in the form of such primordial black holes, define via $\Omega_{PBH}=f\Omega_{CDM}$, such that the photon flux from superradiant neutral pion clouds never exceeds the measured IGRB flux. This representative example yields the upper bound on the primordial black hole abundance $f\lesssim 10^{-7}$ ($7\times10^{-8}$) for a Gaussian (monochromatic) spectrum.

\begin{figure}[htbp]
\includegraphics[scale = 0.45]{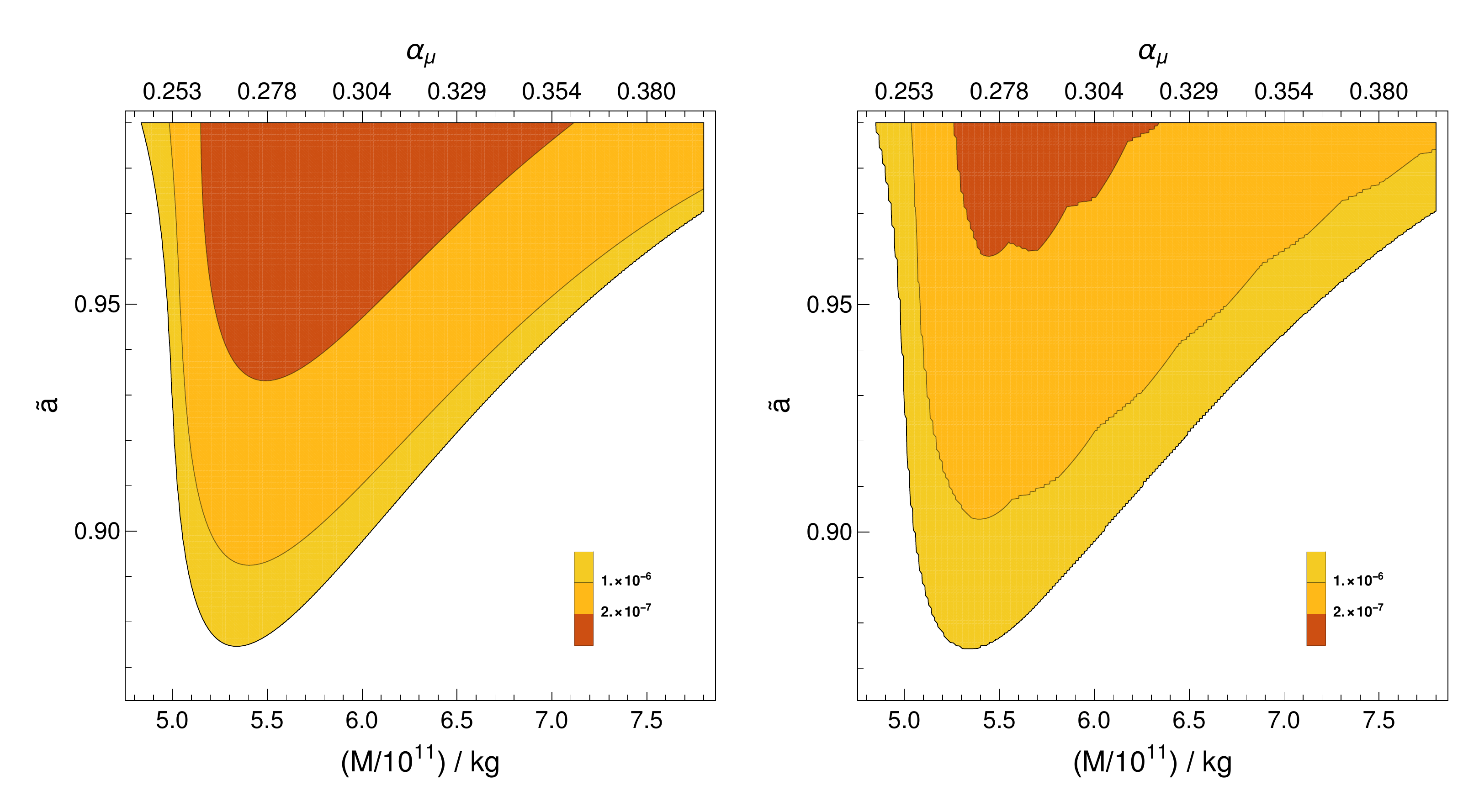}\caption{Upper bounds on the dark matter fraction $f$ in the form primordial black holes of a given mass and spin from neutral pion superradiance, assuming a monochromatic (left) and gaussian (right) photon emission spectrum.}
\label{regionfraction}
\end{figure}

In Fig.~\ref{regionfraction} we show the upper bounds on the primordial black hole abundance in the mass-spin Regge plane, considering only the region where superradiant pion production is effective as previously shown in Fig.~\ref{neutraleffect}.

%For a black hole with such mass and spin, the upper bounds on $f$ agrees with the constraints from evaporation for the same black hole \cite{Carr:2009jm, Arbey:2019vqx}.

The obtained bounds are comparable to those coming from the contribution of Hawking evaporation of such black holes to the IGRB \cite{Carr:2009jm, Arbey:2019vqx}, for high spin values, being slightly more relaxed for the lowest allowed black hole spins. This is not surprising, since for highly spinning black holes a non-negligible fraction of the black hole's mass is converted into gamma-rays via superradiant pion production and subsequent pion decay. Also the Hawking temperature $T_H\sim (8\pi M)^{-1}\sim \mu_0/(8\pi\alpha_\mu)$ is comparable to the neutral pion's mass since $\alpha_{\mu0}\sim \mathcal{O}(0.1)$ in the relevant black hole mass range, so photons from evaporation and superradiant pion decay have comparable energies.

Even though the bounds on $f$ from superradiance and Hawking evaporation are comparable, we note that the photon spectrum is distinct in both cases, which in the future could be used to distinguish these processes in the event of a detection of features in the IGRB spectrum in the few MeV energy range.

%%%%%%%%%%%%%%%%%%%%%%%%%%%%%%%%%%%%%%%%%%%%%%%%%%%%%%%%%%%%%%%%%%%%%%%%%%%%%%%%%%%%%%%%%%
%%%%%%%%%%%%%%%%%%%%%%%%%%%%%%%%%%%%%%%%%%%%%%%%%%%%%%%%%%%%%%%%%%%%%%%%%%%%%%%%%%%%%%%%%%
%%%%%%%%%%%%%%%%%%%%%%%%%%%%%%%%%%%%%%%%%%%%%%%%%%%%%%%%%%%%%%%%%%%%%%%%%%%%%%%%%%%%%%%%%%
%%%%%%%%%%%%%%%%%%%%%%%%%%%%%%%%%%%%%%%%%%%%%%%%%%%%%%%%%%%%%%%%%%%%%%%%%%%%%%%%%%%%%%%%%%

\subsection{Charged pions $\pi^{\pm}$}

\begin{figure}[htbp]
\includegraphics[scale = 0.45]{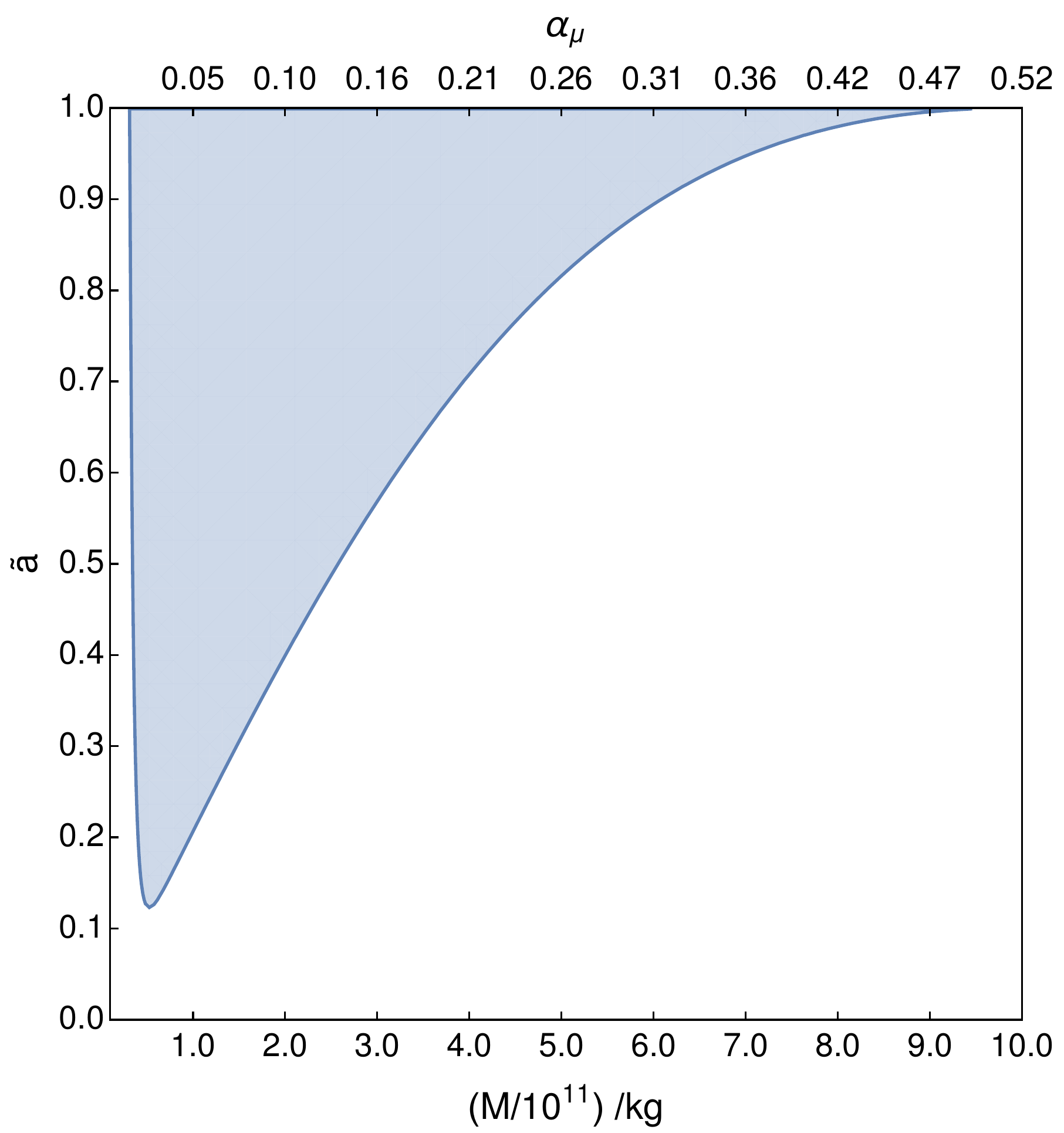}
\caption{Region in the black hole mass-spin Regge plot in which an efficient charged pion production through superradiance is possible (without taking annihilation into account).}
\label{chargeddecay}
\end{figure}

Charged pions have a similar mass to the neutral pions, $\mu_+\simeq 139$ MeV, as a result of the underlying approximate isospin symmetry, but can only decay via weak processes, the dominant one being $\pi^+\rightarrow \mu^+\nu_\mu$ for the positively charged pion. This makes them live much longer than neutral pions, despite the loop-suppression of the electromagnetic neutral pion decay, with a lifetime $\Gamma_{\pi_+}^{-1}\simeq 2.6\times 10^{-8}$ s \cite{Tanabashi:2018oca}. This implies that superradiant charged pion production can be efficient for much smaller black hole spins, as illustrated in Fig.~\ref{chargeddecay}. Self-interactions of the charged pions are also comparable in magnitude to the neutral pion quartic terms, which in principle would allow superradiant $\pi^\pm$ clouds to grow until they are close to the nuclear density. 

\begin{figure}[htbp]
	\centering
	\includegraphics[scale=0.4]{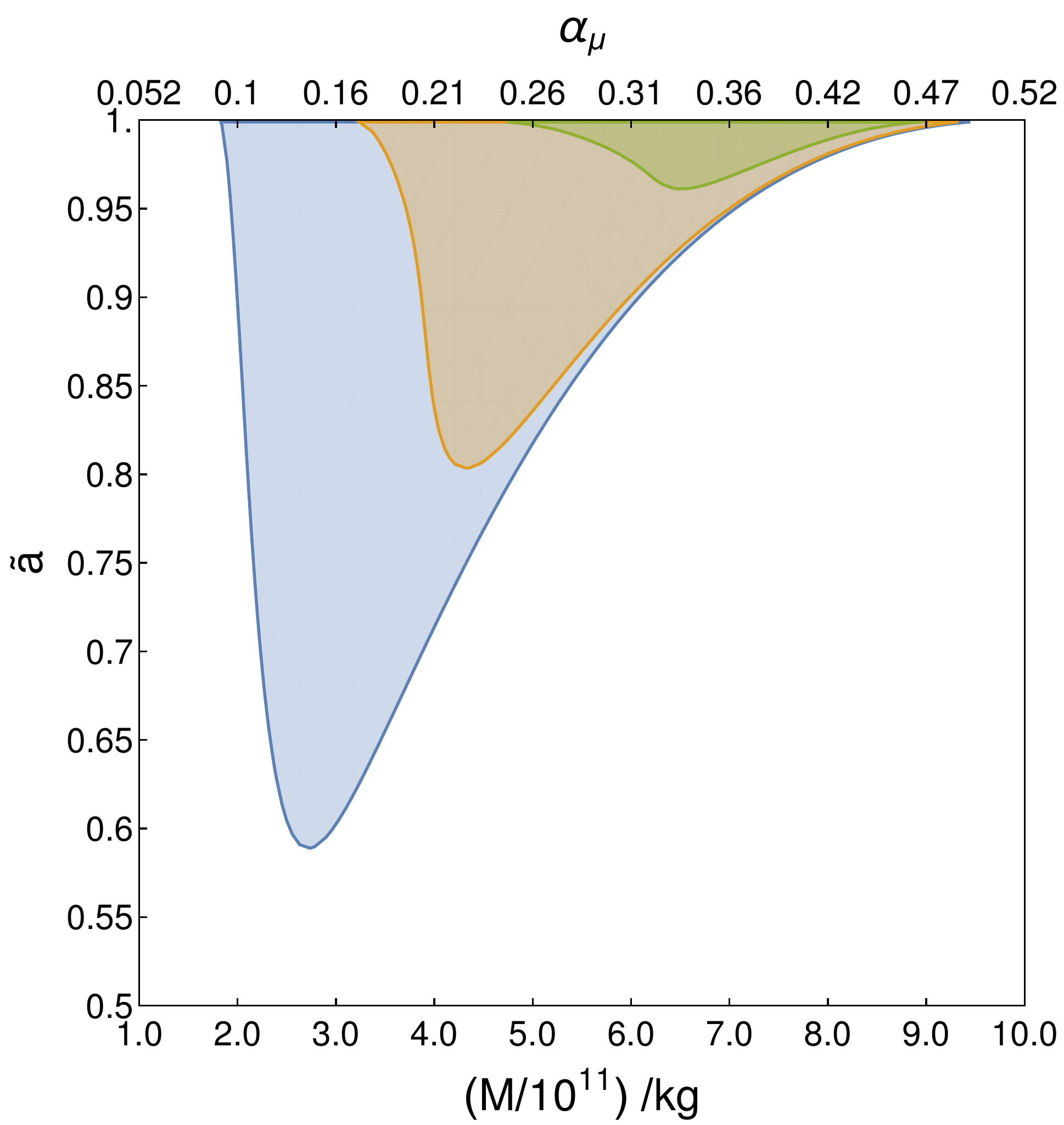}
	\caption{Black hole mass-spin Regge plot with regions for which the critical number of charged pions in the superradiant cloud $N_{c+}>1$ (blue) , $N_{c+}>10$ (brown) and $N_{c+}>50$ (green).}
	\label{critical_annihilation}
\end{figure}

However, one must take into account that other relevant electromagnetic processes, namely charged pion annihilation into photons, $\pi^+\pi^-\rightarrow \gamma\gamma$, may deplete the pion cloud in the charged case. The annihilation rate per pion pair, computed in Appendix A, is given by $\Gamma_a \simeq (\pi \alpha^2)/(2\mu_+^2 V_{cloud})$ in the non-relativistic limit, where $\alpha$ is the fine-structure constant, and $V_{cloud}\simeq 50\pi^2r_0^3$ is the superradiant cloud's volume (i.e.~the volume within which the majority of pions are contained). We also take the number of positively charged pions to be the same as the number of negatively charged pions, since the superradiant instability does not distinguish particles and anti-particles\footnote{Superradiant clouds are presumably born from initial quantum fluctuations in the complex scalar field which may in principle be charged, yielding a final charge asymmetry, but this will not significantly affect our estimates.}. Hence, the number of charged pion pairs in the cloud, $N_+=N_-$, evolves according to
\begin{equation}
	\frac{dN_+}{dt}=(\Gamma_{s+}-\Gamma_{\pi_+})N_+-\Gamma_aN_+^2~,
	\label{chargedNN}
\end{equation}
where we discard self-interactions. There is only a net production of charged pions if $dN_+/dt>0$, yielding an upper bound $N_+<N_{c+}$:
\begin{equation}
	N_{c+} = \frac{\Gamma_{s+}-\Gamma_{\pi_+}}{\Gamma_a}\simeq {\Gamma_{s+}\over \Gamma_a}~.
	\label{Nca}
\end{equation}
We plot this critical number in Fig.~\ref{critical_annihilation}, and one can see that even for large black hole spins we have $N_{c+}\lesssim 50$, such that annihilation quenches the charged pion superradiant instability well below the nuclear density, i.e.~even before self-interactions become relevant for the dynamics. Maintaining a single charged pion pair in the $2p$ bound state requires a significant black hole spin, $\tilde{a}\gtrsim0.6$, and for $\tilde{a}\lesssim 0.8$ superradiance can sustain only a few $\pi^\pm$ pairs in the cloud. We note that for such small occupation numbers our classical treatment of the superradiant instability may break down, but we may confidently conclude that neutral pions are produced in much larger numbers than charged pions within the superradiant clouds.

Using the same approach as in Eq.~(\ref{deltas}), we can estimate how much mass and spin the black hole loses with the condition that the black hole stops producing charged pions effectively when $N_{c+}=0$, i.e.~all pions have annihilated into photons. We may approximate the luminosity of the superradiant cloud due to the photons from charged pion annihilation as $L_+\simeq 2N_{c+}\Gamma_a\mu_+$, where the photon energy is taken to be approximately $E_{\gamma}=\mu_+$. For a black hole with mass $M=5.5\times 10^{11}$ kg and spin $\tilde{a}=0.99$, we estimate that the cloud's lifetime is $t_{cloud}\sim 10^{12}$ yr, much larger than for neutral pions due to the smaller pion number and luminosity. However, since black holes with such mass evaporate in $\sim 10^{10}$ yr, we expect Hawking evaporation to start spinning down the black hole significantly after $\sim1$ Gyr, at which point superradiant charged pion production will also become inefficient compared to annihilation. Hence, taking this into account we do not expect charged pion clouds to live longer than their neutral counterparts. 

Given the much smaller occupation numbers and consequently the lower luminosity, the charged pion clouds will have a much smaller impact on the black hole's mass and spin depletion and on the IGRB flux. We have performed a similar analysis to the one described earlier for the neutral pion clouds, and concluded that the constraints on the primordial black hole abundance are much less stringent in the charged case. For instance, for our working example of a black hole with $M=5.5\times10^{11}$ kg and $\tilde{a}=0.99$, we obtain $f\lesssim 6\times 10^{-5}$ ($3\times 10^{-5}$) for a Gaussian (monochromatic) photon emission spectrum. Thus, if we take into account the effects of both neutral and charged pion production, it is the former that yields the most stringent bounds on a primordial population of highly spinning black holes. Even if charged pion production can occur for much lower black hole spins, the number of charged pairs surviving annihilation is only non-negligible for highly spinning black holes, so that in practice both types of superradiant clouds are only generated for $\tilde{a}\gtrsim 0.8-0.9$.

%%%%%%%%%%%%%%%%%%%%%%%%%%%%%%%%%%%%%%%%%%%%%%%%%%%%%%%%%%%%%%%%%%%%%%%%%%%%%%%%%%%%%%%%%%
%%%%%%%%%%%%%%%%%%%%%%%%%%%%%%%%%%%%%%%%%%%%%%%%%%%%%%%%%%%%%%%%%%%%%%%%%%%%%%%%%%%%%%%%%%
%%%%%%%%%%%%%%%%%%%%%%%%%%%%%%%%%%%%%%%%%%%%%%%%%%%%%%%%%%%%%%%%%%%%%%%%%%%%%%%%%%%%%%%%%%
%%%%%%%%%%%%%%%%%%%%%%%%%%%%%%%%%%%%%%%%%%%%%%%%%%%%%%%%%%%%%%%%%%%%%%%%%%%%%%%%%%%%%%%%%%

\section{Direct detection of superradiant pion clouds}

In addition to considering the effects of a cosmological population of highly spinning primordial black holes on the diffuse IGRB, we may ask whether one can directly detect the photons emitted by superradiant pion clouds within a point-like source containing such primordial black holes. We have concluded earlier that such clouds may live for up to $\sim 1$ Gyr, after which pion production becomes inefficient both in the neutral and charged cases. This implies that we must search for high-redshift sources, $z\gtrsim 7$, which are of course intrinsically difficult to detect. Such sources should also contain a large number of highly spinning primordial black holes so as to maximize the gamma-ray signal, so that we may consider e.g.~looking for super-clusters at high redshifts.

Let us then estimate the expected flux from such a source, namely from neutral pion decay in superradiant clouds, which as we have seen is the dominant process. First, the absolute luminosity of a single primordial black hole that develops a critical superradiant cloud at the nuclear density is, as we have seen before:
\begin{equation}
	L_0= 2\Gamma_{\pi_0}N_{c0}\mu_0\simeq 9\times10^{10}\left({M\over 5\times 10^{11}\ \mathrm{kg}}\right)^{-3}\ \mathrm{W}~.
\end{equation}
An astrophysical object of mass $M_G$, which we assume to be dominated by cold dark matter with a fraction $f$ of highly spinning primordial black holes, will then shine due to neutral pion decay as:
\begin{equation}
	L_0^{G}= f{M_G\over M}L_0\simeq 10^{-4}\left({f\over 10^{-7}}\right)\left({M_G\over M_\odot}\right)\left({M\over 5\times 10^{11}\ \mathrm{kg}}\right)^{-4}L_\odot~,
\end{equation}
where $M_\odot$ and $L_\odot$ are the Sun's mass and total luminosity, respectively. The gamma-ray luminosity of an astrophysical system due to the superradiant pion clouds can thus be quite large, but as we have discussed the objects must be quite far away for superradiant clouds to still be active, so we need to estimate the observable photon flux. The luminosity distance is related to the cosmological redshift via:
\begin{equation}
	d_{L}(z) =cH_0^{-1}\big(z+\frac{1}{2}(1-q_0)z^2+...\big)~,
\end{equation}
where $H_0\simeq 70\ \mathrm{km}\mathrm{s}^{-1}\mathrm{Mpc}^{-1}$ is the present Hubble constant and $q_0\simeq -0.54$ is the present value of the deceleration parameter. This yields $d_{L}(z=7)\sim 190$ Gpc. From this we may compute the flux  $\mathcal{F}_0^G = L_0^G/(4\pi d_L^2)$, obtaining:
\begin{equation}
\mathcal{F}_0^G\simeq 3.5\times 10^{-13}\left({f\over 10^{-7}}\right)\left({M_G\over10^{15} M_\odot}\right)\left({M\over 5\times 10^{11}\ \mathrm{kg}}\right)^{-4}\left({d_L\over200\ \mathrm{Gpc}}\right)^{-2}\ \text{photons cm}^{-2}\text{s}^{-1}~.
\end{equation}
This is, unfortunately, too low a flux to be observed with current technology, even maximizing the primordial black hole dark matter fraction allowed by the IGRB data and considering very large superclusters, such as the protocluster of galaxies observed a redshift $z\simeq 5.7$ with mass $4\times10^{15}M_\odot$ \cite{Jiang:2018hqk}. For comparison, COMPTEL's narrow line sensitivity in the few MeV range is $\sim 10^{-5}-10^{-4}\ \text{photons cm}^{-2}\text{s}^{-1}$, with the proposed AMEGO survey improving this by only one order of magnitude \cite{McEnery:2019tcm}. The estimate above suggests roughly that one would need a detector with an effective area $\sim 10\ \mathrm{m}^2$ operating over a few years to have at least a few photon counts.

If such technology becomes available at some stage in the future, observing a neutral pion decay line may not be sufficient to claim a detection of primordial superradiant clouds. However, a `smoking gun' for primordial pion superradiance could come from observing simultaneously the lines corresponding to neutral pion decay and charged pion annihilation, with a redshift-independent frequency ratio $\mu_0/2\mu_+\simeq 0.486$, and an intensity ratio given by:
\begin{equation}
	\frac{L_{0}}{L_{+}} \simeq {\mu_0\over 2\mu_+}\frac{\Gamma_{\pi_0}}{\Gamma_a}\frac{N_{c0}}{N_{c+}^ 2}\sim 10^4-10^5~.
\end{equation}
Of course the spectral line from charged pion annihilation is much more difficult to detect, requiring even better sensitivity, so this would be extremely hard to probe, but if possible it would provide definite evidence for dense pion clouds in an astrophysical environment, requiring a continuous pion production that compensates for their decay and annihilation, which would strongly suggest primordial black hole superradiance.

One may also speculate about the possibility of keeping superradiance active at later times, for instance if primordial black hole mergers can keep the black holes' spin high, counteracting the effects of both superradiance and evaporation, at least for a fraction of the original population. This would then allow for looking at much closer sources, potentially within the reach of planned detectors such as AMEGO.

%%%%%%%%%%%%%%%%%%%%%%%%%%%%%%%%%%%%%%%%%%%%%%%%%%%%%%%%%%%%%%%%%%%%%%%%%%%%%%%%%%%%%%%%%%
%%%%%%%%%%%%%%%%%%%%%%%%%%%%%%%%%%%%%%%%%%%%%%%%%%%%%%%%%%%%%%%%%%%%%%%%%%%%%%%%%%%%%%%%%%
%%%%%%%%%%%%%%%%%%%%%%%%%%%%%%%%%%%%%%%%%%%%%%%%%%%%%%%%%%%%%%%%%%%%%%%%%%%%%%%%%%%%%%%%%%
%%%%%%%%%%%%%%%%%%%%%%%%%%%%%%%%%%%%%%%%%%%%%%%%%%%%%%%%%%%%%%%%%%%%%%%%%%%%%%%%%%%%%%%%%%

\section{Conclusion}

In this work, we have analyzed the generation of superradiant pion clouds around primordial black holes in the $10^{11}-10^{12}$ kg mass range. We have concluded that an efficient pion production is only possible for very high black hole spins, $\tilde{a}\gtrsim 0.8-0.9$, in order to overcome both the neutral pion decay and charged pion annihilation. In addition, our results show that while neutral pion clouds can reach close to the nuclear density before self-interactions quench superradiance, charged pion clouds have much smaller occupation numbers due to their efficient annihilation into photons.

Photons from neutral pion decays in such superradiant clouds can contribute to the diffuse IGRB. We have computed the resulting photon spectrum, taking into account all the photons emitted from superradiant clouds from the epoch of recombination until superradiance is quenched at $t_{cloud}\sim 1$ Gyr after the Big Bang. We then used the measured IGRB flux in the MeV range to obtain an upper bound $f\lesssim 10^{-7}$ on the fraction of dark matter in primordial black holes in the relevant mass and spin range. This bound is comparable to the IGRB bound from Hawking evaporation in the same mass range, although the latter exhibits only a mild dependence on the black hole spin, while superradiant pion production places bounds exclusively on highly-spinning black holes. Although photons from charged pion annihilation also contribute to the IGRB, the lower density of charged pion clouds makes the resulting flux considerably smaller, yielding much less stringent bounds on the primordial black hole abundance.

Finally, we have considered the possibility of directly detecting both spectral lines associated with superradiant pion clouds, showing that their specific frequency and intensity ratio could constitute a smoking-gun for primordial superradiance. However, since superradiant clouds decay away before the present day, this requires looking at high-redshift sources, resulting in a very small photon flux beyond the reach of current technology. More optimistic direct detection prospects may nevertheless be envisaged in case there is some mechanism to counteract the decrease of the black hole spin due to both superradiance and evaporation beyond the cloud's lifetime estimated in this work, for instance through efficient black hole mergers. This may potentially be significant for primordial black holes living in dense regions, where single or even multiple mergers may increase both the mass and spin of the black holes, possibly bringing some of them into the pion superradiance regime. If this is the case, the corresponding pion gamma-ray lines with the properties estimated in this work could be observable at lower redshifts, and would constitute a smoking-gun for primordial black hole superradiance.

We note that in our analysis we have included the basic processes shaping the dynamics of superradiant pion clouds, namely superradiant pion production and their subsequent electromagnetic decay and/or annihilation. Our conclusions thus rely on these constituting the dominant processes up until pion self-interactions become relevant close to the nuclear density, or equivalently until the pion condensate reaches values near the pion's decay constant within the cloud. Numerical simulations for axion-like fields, which are qualitatively similar to neutral pions, support an onset of field non-linearities and consequent quenching of superradiance when $\phi\lesssim f_\pi$ \cite{Yoshino:2012kn, Yoshino:2015nsa}, while some analyses argue for an earlier saturation of the number density \cite{Arvanitaki:2010sy, Gruzinov:2016hcq}. The latter do not, however, take into account the effect of self-interactions on the mass of the scalar field, as we argue in Appendix B, so we are confident that our estimates constitute a realistic approximation. It should nevertheless be acknowledged that, near the critical density, other nuclear processes may come into play and that these may potentially have some effect on the dynamics of pion clouds and their associated luminosity. Such an analysis is, however, beyond the scope of the present work, where our main goal was to describe the essential dynamical features of these clouds and show that already quite stringent bounds on the abundance of highly spinning primordial black holes can be derived in this context. We nevertheless hope that this work motivates further exploration of this topic.

\acknowledgements

The work of T.\ W.\ K.\ is supported by U.S. DoE grant number DE-SC-0019235. J.\,G.\,R. is supported by the FCT Grant No.~IF/01597/2015, the CFisUC strategic project No.~UID/FIS/04564/2019 and partially by the FCT project PTDC/FIS-OUT/28407/2017 and ENGAGE SKA (POCI-01-0145-FEDER-022217).

 %H2020-MSCA-RISE-2015 Grant No.~StronGrHEP-690904, the FCT project PTDC/FIS-OUT/28407/2017 and the CIDMA Project No.~UID/MAT/04106/2019.

%%%%%%%%%%%%%%%%%%%%%%%%%%%%%%%%%%%%%%%%%%%%%%%%%%%%%%%%%%%%%%%%%%%%%%%%%%%%%%%%%%%%%%%%%%
%%%%%%%%%%%%%%%%%%%%%%%%%%%%%%%%%%%%%%%%%%%%%%%%%%%%%%%%%%%%%%%%%%%%%%%%%%%%%%%%%%%%%%%%%%
%%%%%%%%%%%%%%%%%%%%%%%%%%%%%%%%%%%%%%%%%%%%%%%%%%%%%%%%%%%%%%%%%%%%%%%%%%%%%%%%%%%%%%%%%%
%%%%%%%%%%%%%%%%%%%%%%%%%%%%%%%%%%%%%%%%%%%%%%%%%%%%%%%%%%%%%%%%%%%%%%%%%%%%%%%%%%%%%%%%%%

\appendix

\section{Charged pion annihilation rate}

To compute the annihilation rate of charged pions into photons, $\pi^{+}\pi^{-}\rightarrow \gamma\gamma$, we consider the effective QED Lagrangian for charged pions, which is given by the Lagrangian density:
\begin{align}
	\mathcal{L}=-\partial_{\mu}\phi \partial^{\mu}\phi^{\dagger}-\mu^2\phi\phi^{\dagger}-\frac{1}{4}F^{\mu\nu}F_{\mu\nu}+eA_{\mu}J^{\mu}-e^2A_{\mu}A^{\mu}\phi\phi^{\dagger}\;,
\end{align}
where $J_{\mu}=ie(\phi\partial_{\mu}\phi^{\dagger}-\phi^{\dagger}\partial_{\mu}\phi)$. The corresponding vertices involving charged pions are

\begin{equation*}
	\begin{tikzpicture}[baseline=-\the\dimexpr\fontdimen22\textfont2\relax]
		\begin{feynman}
		\vertex (c);
		\vertex [above left=of c] (f1) {\(\gamma\)};
		\vertex [below left=of c] (i1) {\(\pi^{+}\)};
		\vertex [above right=of c] (f2) {\(\gamma\)};
		\vertex [below right=of c] (i2) {\(\pi^{-}\)};
	
		\diagram*{
			(i1) -- [fermion, edge label'=\(p_1\)] (c) -- [boson, momentum=\(k_1\)] (f1),
			(i2) -- [anti fermion, edge label'=\(p_2\)] (c) -- [boson, momentum=\(k_2\)] (f2),
		};
	\end{feynman}
	\end{tikzpicture}
	= -2ie^2g_{\mu\nu}, \\
	\begin{tikzpicture}[baseline=-\the\dimexpr\fontdimen22\textfont2\relax]
	\begin{feynman}[inline = (v)]
		\vertex [dot] (v);
		\vertex [above left=of v] (i1) {\(\gamma\)};
		\vertex [below left=of v] (i2) {\(\pi^{+}\)};
		\vertex [right=of v] (f) {\(\pi^{+}\)};
		
	\diagram*{
		(i1) -- [boson, reversed momentum=\(k\)] (v) -- [fermion, edge label=\(p_2\)] (f),
		(i2) -- [fermion,edge label'=\(p_1\)] (v),
	};
	\end{feynman} 
\end{tikzpicture}
	=ie(p_1+p_2)^{\mu}
\end{equation*}

The diagrams to compute the matrix element $\mathcal{M}_{fi}$ are, up to order $\mathcal{O}(e^3)$, 	

\begin{equation*}
\begin{tikzpicture}[baseline=-\the\dimexpr\fontdimen22\textfont2\relax]
	\begin{feynman}
		\vertex (c);
		\vertex [right=of c] (d);
		\vertex [above left=of c] (f1) {\(\gamma\)};
		\vertex [below left=of c] (i1) {\(\pi^{+}\)};
		\vertex [above right=of d] (f2) {\(\gamma\)};
		\vertex [below right=of d] (i2) {\(\pi^{-}\)};
	
		\diagram*[horizontal=c to d]{
			(i1) -- [fermion, momentum'=\(p_1\)] (c) -- [boson, momentum'=\(k_1\)] (f1),
			(c) -- [fermion] (d),
			(i2) -- [fermion, momentum=\(p_2\)] (d) -- [boson, momentum=\(k_2\)] (f2),
		};
	\end{feynman}

\end{tikzpicture}
+
(k_1 \rightarrow k_2)
+
\begin{tikzpicture}[baseline=-\the\dimexpr\fontdimen22\textfont2\relax]
	\begin{feynman}
		\vertex (c);
		\vertex [above left=of c] (f1) {\(\gamma\)};
		\vertex [below left=of c] (i1) {\(\pi^{-}\)};
		\vertex [above right=of c] (f2) {\(\gamma\)};
		\vertex [below right=of c] (i2) {\(\pi^{+}\)};
	
		\diagram*{
			(i1) -- [fermion, momentum'=\(p_1\)] (c) -- [boson, momentum=\(k_1\)] (f1),
			(i2) -- [anti fermion, momentum'=\(p_2\)] (c) -- [boson, momentum=\(k_2\)] (f2),
		};
	\end{feynman}

\end{tikzpicture}
\end{equation*}

This results in the matrix element $\mathcal{M} = -ie^2\epsilon_{\mu}(k_1)\epsilon_{\nu}(k_2)\mathcal{M}^{\mu\nu}$, with
\begin{align}
	\mathcal{M}^{\mu\nu}=2\Bigg[2\frac{p_1^{\mu}(k_1-p_1)^{\nu}}{\mu^2-t}+2\frac{p_1^{\nu}(k_2-p_1)^{\mu}}{\mu^2-u}		+g^{\mu\nu}\Bigg]\;,
\end{align}
where we used the fact that the Ward-Takashi identities are satisfied, i.e. $k_{1\mu}M^{\mu\nu}=k_{2\mu}M^{\mu\nu}=0$ and that $\epsilon(k_i)\cdot k_i=0$.
After averaging over the photon polarizations and some algebra, we obtain
\begin{align}
	\langle|\mathcal{M}|^2\rangle_{pol}=2e^4\frac{5\mu^8-4(u+t)\mu^6+\mu^4(u^2+t^2)+u^2t^2}{(\mu^2-t)^2(\mu^2-u)^2}\;.
\end{align}
In the non-relativistic regime, $s\simeq 4\mu^2$. Using the equality $s+t+u=2\mu^2$ and $t\simeq -\mu^2$, it leads to 
\begin{align}
	\langle|\mathcal{M}|^2\rangle_{pol} \simeq 2e^4 = 32\pi^2\alpha^2,
\end{align}
where $\alpha$ is the fine structure constant.

For such a configuration of particle scattering, $\pi^{-}(p_1)+\pi^{+}(p_2)\rightarrow\gamma(k_1)+\gamma(k_2)$, the cross section takes the form:
\begin{align}
	\sigma_i = \frac{(2\pi)^4}{4E_{p_1}E_{p_2}|v_1-v_2|}\int\frac{d^3k_1}{(2\pi)^32E_{k_1}}\frac{d^3k_2}{(2\pi)^32E_{k_2}}\delta(p_1+p_2-k_1-k_2)|\mathcal{M}_{fi}|^2\;,
\end{align}
which in the center-of-mass frame can be simplified to
\begin{align}
	\sigma_i =\frac{|\mathcal{M}_{fi}|^2}{16\pi s}\frac{|\vec{k}_1|}{|\vec{p}_1|}= \frac{|\mathcal{M}_{fi}|^2}{32\pi\sqrt{s(s-4\mu^2)}}\;.
\end{align}

The annihilation rate is computed by multiplying the incident flux and the cross section. Considering that we have the same number density for each species of incident particles, then the flux is given by $f = n_{\pi^+}|v_1-v_2|$, resulting in the annihilation rate per particle:
\begin{align}
	\Gamma_a &=\frac{2}{V_{cloud}} \frac{32\pi^2\alpha^2}{32\pi s}\\
	&\simeq \frac{\pi\alpha^2}{2\mu^2V_{cloud}}~.
\end{align}

%%%%%%%%%%%%%%%%%%%%%%%%%%%%%%%%%%%%%%%%%%%%%%%%%%%%%%%%%%%%%%%%%%%%%%%%%%%%%%%%%%%%%%%%%%
%%%%%%%%%%%%%%%%%%%%%%%%%%%%%%%%%%%%%%%%%%%%%%%%%%%%%%%%%%%%%%%%%%%%%%%%%%%%%%%%%%%%%%%%%%
%%%%%%%%%%%%%%%%%%%%%%%%%%%%%%%%%%%%%%%%%%%%%%%%%%%%%%%%%%%%%%%%%%%%%%%%%%%%%%%%%%%%%%%%%%
%%%%%%%%%%%%%%%%%%%%%%%%%%%%%%%%%%%%%%%%%%%%%%%%%%%%%%%%%%%%%%%%%%%%%%%%%%%%%%%%%%%%%%%%%%

\section{On the effects of non-linear scalar field interactions}

It has been claimed in the literature \cite{Arvanitaki:2010sy, Gruzinov:2016hcq} that scalar self-interactions for pseudo-scalar fields such as axions or axion-like fields like pions become relevant for field values within the cloud below the corresponding decay constant. Such a claim is based on a non-relativistic approach to the problem, which we argue here must include the effects of self-interactions on the field's mass, being otherwise inconsistent.

Let us consider the Klein-Gordon equation for a real scalar field with quartic self-interactions, representing the leading non-linear effect in the scalar potential:
\begin{equation}
-g^{\mu\nu}\nabla_\mu\nabla_\nu \phi + \mu^2\phi -{\lambda\over 6}\phi^3 =0~.
\end{equation}
 Let us consider the non-relativistic/weak gravity limit of this equation in a Schwarzschild background (rotation is irrelevant for this discussion). For this, we consider an {\it ansatz} for the field:
\begin{equation}
\phi = {1\over \sqrt{2\omega}}\left(\psi e^{-i\omega t} + \psi^*e^{i\omega t}\right)~,
\end{equation}
where we assume that the spatial and temporal variation of $\psi$ occurs on scales much larger than $\omega^{-1}$. For $r\gg 2M$ (justified in the non-relativistic limit $\alpha_\mu \lesssim 1$), the equation for $\psi$ is given by:
\begin{equation}
\left[\ddot\psi -2i\omega \dot\psi -\left(1+{2M\over r}\right)\omega^2 \psi -\nabla^2\psi + \mu^2\psi -{\lambda\over 4\omega}|\psi|^2\psi\right]e^{-i\omega t} + \mathrm{c.c.} = {\lambda\over 12\omega }\psi^3 e^{-3i\omega t} + \mathrm{c.c.} 
\end{equation}
Averaging over the rapidly oscillating terms and taking $\ddot\psi \ll 2i\omega \dot\psi$ in the non-relativistic approximation where $\psi$ varies slowly compared to $e^{\pm i\omega t}$, we obtain the Schr\"odinger-like equation:
\begin{equation}
 i \dot\psi =  -{\nabla^2\psi\over 2\omega} +{1\over2\omega}\left[-\left(1+{2M\over r}\right)\omega^2   + \mu^2 -{\lambda\over 4\omega}|\psi|^2\right]\psi~.
 \end{equation}
Now, to leading order we can neglect the spatial and temporal gradients of $\psi$ (and consistently the gravitational effects), yielding:
\begin{equation} \label{omega}
\omega^2 = \mu^2 -{\lambda\over 4\omega}|\psi|^2~.
 \end{equation}
Hence, in the absence of the non-linear potential term we may consistently set $\omega=\mu$ as standard in the literature. However, we see that the non-linear term contributes to the effective mass of the scalar field, and for constant $\psi$ this would be the leading non-linear effect. Field gradients prevent us from absorbing all the non-linear effects into the effective mass, but we may use e.g.~a mean-field approach and split the non-linear potential term as follows:
\begin{equation}
 i \dot\psi =  -{\nabla^2\psi\over 2\omega} - {M\omega\over r} \psi -{\lambda\over 8\omega^2}\left(|\psi|^2-|\psi_0|^2\right)\psi +{1\over2\omega}\left(-\omega^2   + \mu^2 -{\lambda\over 4\omega}|\psi_0|^2\right)\psi~.
 \end{equation}
where the average particle number density is given by:
\begin{equation}
 |\psi_0|^2 = \langle |\psi|^2\rangle = {1\over N} \int d^3x |\psi|^4~, 
 \end{equation}
with $N=\int d^3x |\psi|^2$ denoting the total particle number within the scalar cloud. 
This then consistently allows for the inclusion of the mean-field effects into the field's effective mass $\omega^2=\mu^2- \lambda |\psi_0|^2/4\omega$, leaving the non-relativistic Schr\"odinger equation:
\begin{equation}
 i \dot\psi =  -{\nabla^2\psi\over 2\omega} - {M\omega\over r} \psi -{\lambda\over 8\omega^2}\left(|\psi|^2-|\psi_0|^2\right)\psi~. 
  \end{equation}
Now, if we take the non-linear term as a perturbation to the gravitational problem, we see that:
\begin{equation}
\langle V_{NL} \rangle =-{\lambda\over 8\omega^2} {1\over N}\int d^3 x (|\psi|^2-|\psi_0|^2)|\psi|^2 =0~,
  \end{equation}
so that this term does not modify the spectrum of gravitationally bound states to leading order in perturbation theory, independently of the average particle number density, and effects arise only at $\mathcal{O}(\lambda^2)$. Note also that the scattering cross-section for $\phi\phi\rightarrow \phi\phi$ is also an $\mathcal{O}(\lambda^2)$ effect.   

Thus, considering simply the non-linear self-interactions obtained in the non-relativistic regime discarding the mean-field contribution to the effective scalar mass in the presence of a non-trivial field  condensate is, as shown above, an inconsistent procedure. The non-relativistic effects of the self-interactions are much smaller than their contribution to the field's mass, and in fact when $\phi\sim f_\pi$ (i.e.~$|\psi_0|^2\sim \mu^3/\lambda $) the effective mass $\omega$ becomes too small for the non-relativistic approximation to hold.

\end{document}